# Search for very low-mass objects in the Galactic Halo *

The EROS collaboration

E. Aubourg[1], P. Bareyre[1], S. Bréhin[1], M. Gros[1], J. de Kat[1], M. Lachièze-Rey[1], B. Laurent[1], E. Lesquoy[1], C. Magneville[1], A. Milsztajn[1], L. Moscoso[1], F. Queinnec[1], C. Renault[1], J. Rich[1], M. Spiro[1], L. Vigroux[1], S. Zylberajch[1], R. Ansari[2], F. Cavalier[2], M. Moniez[2], J.-P. Beaulieu[3], R. Ferlet[3], Ph. Grison[3], A. Vidal-Madjar[3], J. Guibert[4], O. Moreau[4], F. Tajahmady[4], E. Maurice[5], L. Prévôt[5] and C. Gry[6]

[1] CEA, DSM, DAPNIA, Centre d'Etudes de Saclay, 91191 Gif-sur-Yvette, France
[2] Laboratoire de l'Accélérateur Linéaire, Centre d'Orsay, 91405 Orsay Cedex, France
[3] Institut d'Astrophysique de Paris, 98 bis Boulevard Arago, 75014 Paris, France
[4] Centre d'Analyse des Images de l'I.N.S.U, Observatoire de Paris, 61 avenue de l'Observatoire, 75014 Paris, France
[5] Observatoire de Marseille, 2, place Le Verrier, 13248 Marseille Cedex 04, France
[6] Laboratoire d'Astronomie Spatiale de Marseille, Traverse du Siphon, Les Trois Lucs, 13120 Marseille, France

Received date; accepted date

**Abstract.** We present results from a search for gravitational microlensing of stars in the Large Magellanic Cloud by low mass objects in the Galactic Halo. The search uses the CCD light curves of about 82,000 stars with up to 46 measurements per night over a period of 10 months. No light curve exhibits a form that is consistent with a microlensing event of maximum amplification greater than 1.2. This null result makes it unlikely that the Halo is dominated by objects in the mass range $5\ 10^{-8} M_\odot < M < 5\ 10^{-4} M_\odot$.

The presence of large quantities of "dark matter" in spiral galaxies like our own has been inferred from their flat rotation curves (Primack et al. 1988). Though a variety of new weakly interacting elementary particles have been proposed to make up the dark matter, compact objects are also viable candidates if they are in a sufficiently dim form. A possible form would be objects too light to burn hydrogen ($M < 0.07 - 0.1 M_\odot$) (Carr 1990).

We report here results from a search for unseen compact objects in the Galactic Halo being performed by our collaboration "EROS" (Expérience de Recherche d'Objets Sombres) at the European Southern Observatory at La Silla, Chile. Such objects can be detected via the gravitational microlensing effect (Paczyński 1986) which would lead to an apparent temporary brightening of stars outside our Galaxy as the unseen object passes near the line of sight. The amplification is given by $A = (u^2+2)/[u(u^2+4)^{1/2}]$ where $u$ is the undeflected "impact parameter" of the light ray with respect to the unseen object in units of the "Einstein Radius", $R_E = (4GM_d Lx(1-x)/c^2)^{1/2}$.

Here, $M_d$ is the deflector mass, $L$ is the observer-source distance, and $Lx$ is the observer-deflector distance. EROS monitors stars in the Large Magellanic Cloud (LMC) with $L=$ 55 kpc yielding typical values $R_E \sim 2\ 10^3 R_\odot \sqrt{M_d/M_\odot}$. For the "standard" spherical isothermal Halo model (Primack et al. 1988), the rate per monitored star for microlensing events with amplifications greater than a threshold amplification $A_T$ corresponding to an impact parameter $u_T$ has been calculated to be $1.66\ 10^{-6}\ u_T\ (M_\odot/M_d)^{1/2}$ yr$^{-1}$ (Griest 1991). The model assumes a total Galactic "dark" mass of $4.0\ 10^{11} M_\odot$ within 50 kpc of the galactic center yielding a flat rotation curve out to the position of the LMC. The time scale for the amplification is the time for a Halo object to move through an angle corresponding to its Einstein radius and its average is $\tau \sim 75$ days$\sqrt{M_d/M_\odot}$. The resulting achromatic light curve has a characteristic shape and, given the small rate for microlensing and the preponderance of intrinsically stable stars, the event should be the only significant variation on the curve. Because of geometry, the events are uniformly distributed in impact parameter at maximum amplification, yielding a distribution of maximum amplifications that falls rather slowly with increasing amplification ($dN/dA \propto 1/A^2$ for $A \gg 1$).

We have previously reported results from our Schmidt-plate search for long time-scale microlensing ($\tau > 2$ days) (Aubourg et al. 1993). Two light curves were found that were consistent with the microlensing hypothesis. Candidate events have been reported by the MACHO collaboration (Alcock et al. 1993, Bennett 1994). The OGLE collaboration (Udalski et al. 1993) has reported the observation of the microlensing of stars in the Galactic Bulge.

Here, we report on results (described in detail in (Queinnec 1994) ) from our search for short time-scale ($\tau < 7$ days) microlensing. It uses a CCD camera to monitor stars in one field ($1.1° \times 0.4°$) of the LMC bar. The exposure time was typically 10 minutes with up to 46 alternating red and blue images taken per night. The program is therefore sensitive to microlensing events with time scales larger than 30 minutes. About 45000

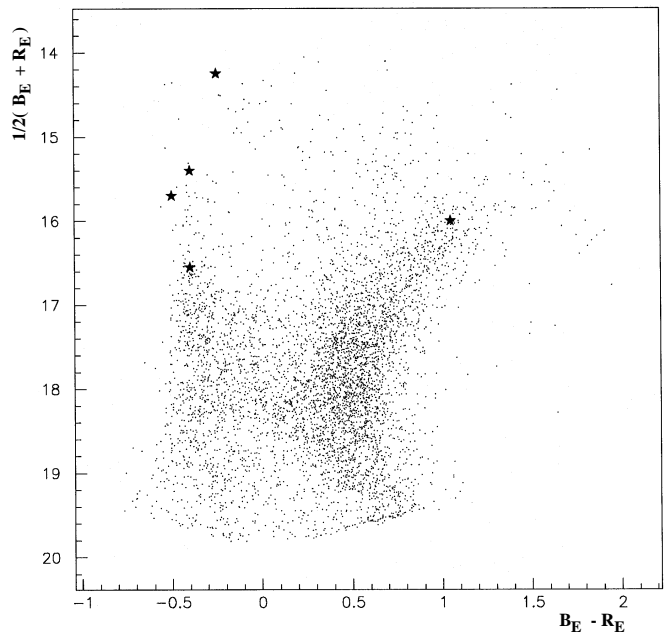

**Fig. 1.** A sample colour-magnitude diagram. The magnitudes $B_E$ and $R_E$ correspond to EROS blue and red filters (Arnaud 1994); these magnitudes have not yet been related to standard photometric systems. In the present search for microlensing events only the amplification (and thus the magnitude change) is relevant. The stars corresponding to the last five events (see text) are shown as filled stars. They are all much brighter than the average.

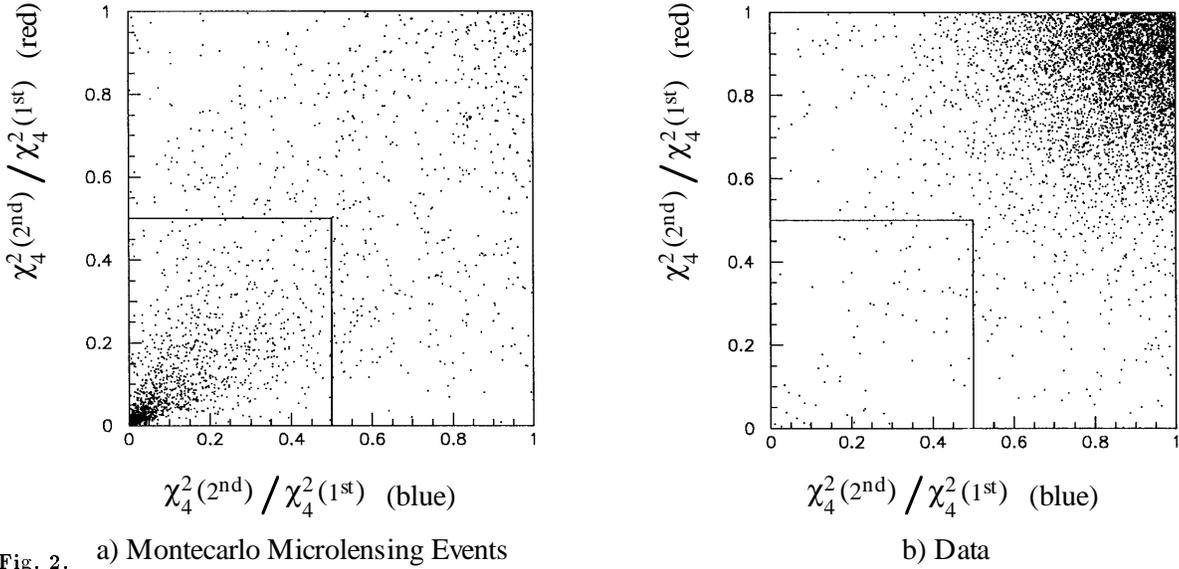

**Fig. 2.** a) Montecarlo Microlensing Events   b) Data

The ratio of the $\chi_4^2$'s of the second and first most significant variations for the red light curves vs. that for the blue. Fig. 2a shows the Monte Carlo generated events for $M_d = 10^{-6} M_\odot$. The accumulation of events in the lower left is due to events with one and only one significant variation, as expected for microlensing of stable stars. (Figures obtained for other values of $M_d$ are very similar to fig. 2a.) Fig. 2b shows the data. Most events are in the upper right as expected for intrinsically stable stars with no significant variations and for periodic variable stars. The position of the cuts are shown in both figures.

March 31, 1992 while about 82000 stars were monitored between August 21, 1992 and March 31, 1993 and again for the same period in 1993 to 1994. This paper presents the first results from this program using 1800 measurements from 1991-92 and 4000 measurements from 1992-93. Using the calculated rate (Griest 1991), the total number of microlensing events of amplification greater than 1.3 expected to occur during the observing time is about $0.1(M_\odot/M_d)^{1/2}$, if the Halo is fully comprised of objects of mass $M_d$. We can therefore hope to observe one or more events if the dominant component of the Halo consists of objects in the range $\sim 10^{-8} M_\odot < M_d < 10^{-2} M_\odot$. (Below this range the event durations are often less than the exposure times.)

A full description of the CCD camera and telescope can be found in (Arnaud et al. 1994). The camera consists of 16 buttable $579 \times 400$ pixel Thomson THX 31157 CCDs. Eleven CCD's were active in 1991-92 and 15 in 1992-93. The camera is mounted on a 40 cm reflector (F/10) refurbished by us and the Observatoire de Haute-Provence.

The first step of the analysis procedure was to construct one reference image for each colour and for each season by combining 50 images taken with good atmospheric conditions. (Because of small differences in the filters used and field observed, the 1991-92 and 1992-93 data have been treated separately but with the same programs.) From the reference images, a star finding algorithm then established a star catalogue. After elimination of stars near defective pixels, a final star catalogue for each season was then formed by matching stars in the two colours. The final catalogue then contains about 82000 (45000) stars for the 1992-93 (1991-92) season. The limiting magnitude corresponds to $m_V \sim 19$ with about 30% main sequence and 70% giants. A sample colour-magnitude diagram from 1 CCD is shown in figure 1.

After the establishment of the star catalogue, light curves were constructed by treating individual images. First, each image was geometrically aligned with the reference using bright isolated stars. The star positions in the reference image then served as input to a photometric fitting program to determine the luminous flux of each catalogue star on the new image. Successive images then add one point to the blue or red light curve of each star in the catalogue. Photometric errors associated with each point on the curve are estimated empirically from the point-to-point variations on a given curve and from the overall image quality. They are typically 6 % r.m.s.

After the elimination of images of poor quality, each light curve is subjected to a series of cuts chosen to isolate microlensing-like events. As explained below, the efficiency of these cuts to accept real microlensing events is determined by applying the same cuts to Monte Carlo microlensing events that are constructed by amplifying points on randomly selected experimental light curves.

Because of the large volume of data, the first series of cuts uses only quantities that can be rapidly calculated. For each set of four neighboring measurements in a given colour, we calculated a quantity, $\chi_4^2$, related to the deviation of the measured fluxes from the reference flux, $\phi_{ref}$, defined as the most probable value on the light curve :

$$\chi_4^2 = \sum_{i=1}^{4} \left( \frac{\phi_i - \phi_{ref}}{\sigma_i} \right)^2,$$

where $\phi_i$ is the flux associated with the point $i$ and $\sigma_i$ is its estimated uncertainty. The cuts use the $\chi_4^2$ for the first and second most significant variations in each colour. Additionally, we use the quantity $\bar{\chi}^2$ calculated as in the above formula except that the sum runs over all points not near the most significant variation.

The great majority of stars exhibit only random fluctuations due to measurement errors. These stars are mostly eliminated by the loose requirement that the most significant variation in the blue be within 15 days of the most significant variation in the red. Intrinsically variable stars with a very significant second variation are eliminated by requiring that $\chi_4^2 < 80$ for the second variation in each colour. Variable stars with long term variations are eliminated by requiring in each colour that the $\bar{\chi}^2$ be less than 2.5 times the number of points.

After these very loose cuts we are left with about 15 % of the original light curves. Because the errors on individual photometric measurements are determined only approximately, our next cut uses only the ratio of the $\chi_4^2$ values for the most significant and the second most significant variations. Figure 2 shows the ratio of the $\chi_4^2$ values of the second and first most significant variations for the red vs. the same ratio for the blue. The Monte Carlo generated events (Fig. 2a) are accumulated in the lower left because they have one and only one significant variation, as expected for microlensing of stable stars. Fig. 2b shows the data. Most events show comparable first and second variations as expected for stable stars with variations coming only from measurement errors. Requiring that the ratio be less than 0.5 in both colours leaves us with only 88 stars. Their light curves are then examined in detail and fitted for the theoretical microlensing light curve, neglecting possible star size effects.

Most of the 88 stars show an "unphysical" discontinuous flux variation, generally due to inaccurate photometry due to bad atmospheric conditions or inaccurate telescope guiding. These stars are eliminated by requiring a good agreement between the time of maximum variation in the red, $t_R$, and that in the blue, $t_B$. Specifically, we require $t_B - t_R < 4\sigma_t$ where $\sigma_t = 0.05$ day $\sqrt{\tau/(1 \text{ day})}$ is the mean uncertainty in the time of maximum of the light curve. After this cut 11 stars remain.

Six of the remaining stars have variations on long timescales ($\tau > 7$ days) and are concentrated in regions of the colour-magnitude diagram known to contain many variable stars. These curves will be discussed in a later publication. For the purposes of this paper on short time-scale microlensing, we make a cut requiring $\tau < 7$ days leaving us with five stars. This significantly reduces our efficiency for microlensing events only if the lensing objects have $M_d > 10^{-3} M_\odot$.

The five remaining stars are indicated in fig. 1; they show very small flux variations, of an amplitude comparable with the photometric resolution. All events have reconstructed amplifications less than 1.16 which, if they were indeed microlensing events, would correspond to impact parameters, $u > 1.4$. Figure 3 shows the distribution of fitted impact parameters, $u$, for Monte Carlo events and for the five observed events. In contrast to the observed events, the expected distribution for microlensing events is concentrated at small impact parameters. We therefore make a final cut requiring impact parameters $u < 1.3$ leaving no candidates.

The efficiency of the cuts to accept real microlensing events is estimated with Monte Carlo generated lensing events, superimposed on a random sample of the experimental light curves.

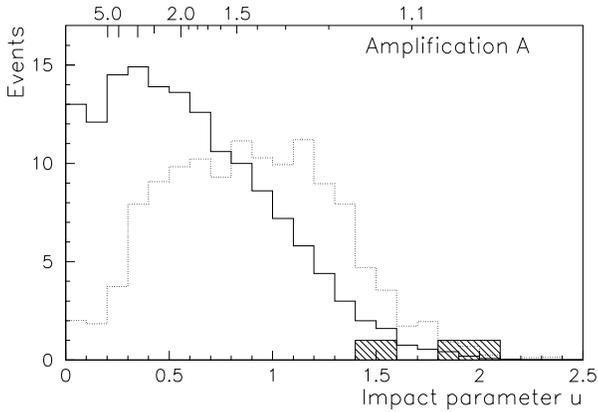

**Fig. 3.** The expected distributions of reconstructed impact parameters, $u$, for $M_d = 10^{-5} M_\odot$ (solid line) and $M_d = 10^{-7} M_\odot$ (dashed line). The normalisation of the two distributions is arbitrary, but their relative normalisation is correct. The dip observed at small $u$ (large amplification) for the distribution at $M_d = 10^{-7} M_\odot$ is due to the fact that we fit the theoretical light curve for negligible star size, while the Monte Carlo events are generated taking into account the actual radius of the source stars. This only results in an overestimation of the fitted impact parameter $u$ but has a small effect on the detection efficiency. Also shown are the five observed events (hatched area). They are concentrated at higher impact parameters (low amplifications).

The generated events follow the flat geometric distribution of impact parameters, $u$. The relation between $u$ and amplification was modified to take into account two effects. First, the finite size of the observed star means that all points on the star will not be amplified by the same factor. This effect is important only for $M_d < 10^{-6} M_\odot$ where $R_E$ is less than the typical stellar radius. The calculated number of expected events is reduced by 25 percent for $M_d = 10^{-7} M_\odot$. Second, in these very dense star fields, stars may be "blended" so that a light curve may receive significant contributions from more than one star. While this means that we effectively monitor more than one star with each light curve, the amplification of a star by a given amount $A_{real}$ will be reconstructed as a smaller amplification $A_{rec}$ by the photometric programs. This effect has been estimated by treating Monte Carlo fabricated images that use as input the measured star population in the LMC down to luminosities a factor 10 dimmer than the dimmest reconstructed by EROS. For our sample of stars, it was found that $(A_{rec} - 1) \sim \beta(A_{real} - 1)$ with $\beta = 0.75$ for the brightest star associated with the light curve and $\beta = 0.15$ for the second brightest star associated with the light curve. The overall effect is to reduce the number expected by about 8 (20) percent for $M_d = 10^{-6} M_\odot$ ($10^{-7} M_\odot$).

Table 1 shows the expected number of events as a function of the deflector mass for a standard spherical isothermal Halo comprised only of objects of that mass. The expected number of events is greater than 2.3 for $5\ 10^{-8} < M_d/M_\odot < 7\ 10^{-4}$ so we exclude this mass range at the 90% C.L. under the assumption that all objects in the Halo have the same mass. The expected number of events is greater than 6.9 for $3\ 10^{-7} < M_d/M_\odot < 1.5\ 10^{-5}$ so in this mass range we exclude the possibility that such objects could account for as much as one third of the Halo. The excluded range applies to any distribution of mass that is sufficiently concentrated in the above range. For example, we consider a deflector mass distribution of the form

$$\frac{dN}{dM} \propto M^{-\alpha} \qquad M_{min} < M < 0.07 M_\odot$$

and $dN/dM = 0$ otherwise. Figure 4 shows the excluded zone of the parameter space $(\alpha, M_{min})$. For $\alpha > 2$ the Halo mass is dominated by objects of mass near $M_{min}$ and we rule out, for $\alpha > 3$, the range $5\ 10^{-8} < M_{min}/M_\odot < 5\ 10^{-4}$. Near $\alpha = 2$ where each decade of mass contains the same total mass, the region $10^{-12} < M_{min}/M_\odot < 10^{-5}$ is ruled out. For $\alpha < 2$ the Halo mass is dominated by high-mass objects and we derive no interesting limits.

**Table 1.** The expected number of microlensing events as a function of deflector mass, for a standard spherical isothermal Halo comprised only of objects of that mass.

| $M_d/M_\odot$ | $10^{-3}$ | $10^{-4}$ | $10^{-5}$ | $10^{-6}$ | $10^{-7}$ |
|---|---|---|---|---|---|
| number of events | 1.9 | 4.3 | 7.5 | 9.7 | 5.6 |

The numbers in Table 1 have been obtained for the assumption of a spherical Halo. The precise mass limits depend on the assumed phase space distribution of lenses. Using instead a flattened Halo (down to $c/a = 1/3$) does not change these numbers by more than 20 percent. Other possibilities are discussed e.g. in (Sackett 1993, Giudice 1993, Gould 1994, Frieman 1994, Evans 1994).

In summary, we have searched for microlensing events with time scales ranging from 30 minutes to 7 days. The lack of candidates in this range places significant constraints on any model of the Halo that relies on objects in the range $5\ 10^{-8} < M/M_\odot < 5\ 10^{-4}$. We note that hydrogenous objects of masses below this range would have been expected to evaporate before the present epoch (de Rújula et al. 1992). We are continuing to take and analyze data and expect to improve these results soon.

*Acknowledgements.* We thank C. Alcock, D. Bennett, A. Bijaoui, and C. Stubbs for discussions. We thank Ph. Gitton, J.-P. Kneib, and the staff of the European Southern Observatory for help with the observations.

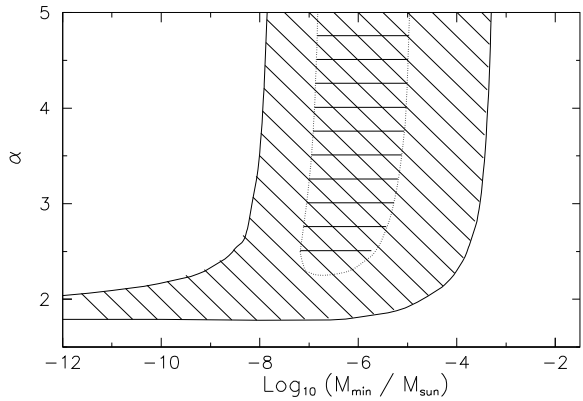

**Fig. 4.** The parameter space $(\alpha, M_{min})$ as explained in the text. Inside the outer contour the number of events expected would be greater than 2.3 and is excluded at 90% C.L. Inside the inner contour the number of events expected would be greater than 6.9 and we limit compact objects to less than 1/3 of the total Halo.